# A combined field approach for the two-way coupling problem in the liquid evaporation

*Xuefeng Xu*[*]

School of Technology, Beijing Forestry University, Beijing 100083, China

**ABSTRACT**: During liquid evaporation, the temperature of the liquid determines the saturated vapor pressure above it, which controls the evaporation rate and thus determines the liquid temperature through latent heat. Therefore, the equations for the vapor concentration in the atmosphere and for the temperature in the liquid are coupled and must be solved in an iterative manner. In the present paper, a combined field approach which unifies the coupled fields into one single field and thus makes the iteration unnecessary is proposed. The present work will be useful in scientific and industrial processes involving liquid evaporation and may also have more general applications to coupled field problems in which all the fields have the same governing equation.

---

[*] Corresponding author. E-mail: xuxuefeng@bjfu.edu.cn,.

# 1. Introduction

The evaporation of liquid is not only a common phenomenon in daily life, but also a fundamental process that impacts a wide range of industrial and scientific applications. For example, a sessile liquid droplet often leaves the solute particles on the substrate, forming different patterns of deposition upon drying.[1-5] This phenomenon has been used as the basis of many applications including DNA-RNA mapping[6,7] and ink-jet printing of functional materials.[8] Controlling the distribution of the particle deposition after the liquid has dried is vital in these applications. This needs a better understanding for the process of the liquid evaporation.

For the case of the steady-state diffusion-controlled evaporation, an exact solution for the evaporation flux of pinned sessile liquid droplets was derived by Picknett and Bexon[9] and by Deegan et al.[2] by simply assuming that the atmosphere just above the liquid surface is saturated with vapor and that the saturation concentration of vapor is a constant along the liquid surface and by using the known solution of an equivalent electrostatic problem.[10] Further, a simple approximate expression for the evaporation flux along the droplet surface was obtained numerically by Hu and Larson.[11]

By allowing the saturation concentration of vapor just above the droplet surface to be a function of the liquid temperature, the theoretical model of the droplet evaporation was generalized to include the effect of the evaporative cooling by Dunn et al.,[12,13] Sefiane et al.,[14] and Saada et al.[15]. In the generalized model, the evaporation lowers the temperature of the liquid at the droplet surface, which in turn alters the saturation concentration of vapor there.[12-16] This implies that the problem of the vapor concentration in the air is coupled with the problem of the temperature in the droplet and thus must be solved numerically. During the computation procedure, the iteration between these two coupled physics fields is needed.

To investigate the evaporative cooling effect in the liquid evaporation, an efficient way to compute directly the vapor concentration field and the temperature field without iteration can be very helpful. For this goal, a combined field approach is introduced for the evaporation of liquid droplet in the present paper. This approach can combine the coupled physics fields into one single field and thus make the iteration unnecessary. The present study may contribute to the studies of the liquid evaporation and thus may be useful to control the flow and the deposition of drying droplets.

**2. Mathematical Model**

Here, we consider a small, pinned, and slowly evaporating liquid droplet with contact angle of $\theta$ and contact line radius of $R$ resting on a flat isothermal substrate with constant temperature $T_0$ which is equal to the room temperature. For the small and slowly evaporating droplet, the droplet shape is regarded as a spherical cap due to its small Bond number and capillary number.[11,16] Due to the axisymmetric configuration, a cylindrical coordinate system ($r$, $z$) is chosen (Figure 1).

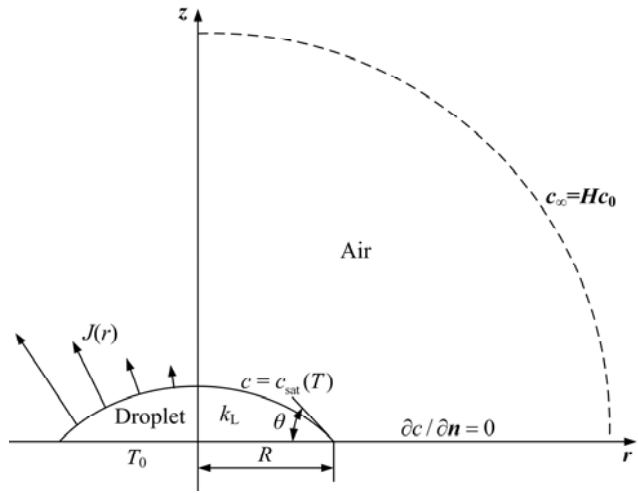

**Fig. 1** A spherical-cap evaporating liquid droplet on a flat substrate in a cylindrical coordinate system with radial coordinate $r$ and axial coordinate $z$.

During the evaporation of the small and slowly evaporating droplets, it is reasonable to assume that both the temperature $T$ in the liquid and the vapor concentration $c$ in the atmosphere satisfy Laplace's equation, i.e., $\nabla^2 T = 0$ and $\nabla^2 c = 0$.[1-3,11,16-19] At the liquid-vapor interface, the vapor concentration $c$ is assumed to be the saturation concentration which is assumed to be a linear increasing function of the temperature, namely,

$$c = c_{sat}(T) = c_0 + b(T - T_0) \tag{1}$$

where $T$ is the local liquid temperature at the droplet surface, $b = \left.\dfrac{dc_{sat}(T)}{dT}\right|_{T=T_0}$, and $c_0 = c_{sat}(T_0)$ is the saturated vapor concentration of the liquid at temperature $T_0$.[12,13] $c_\infty = Hc_0$ is the vapor concentration far above the droplet, where $H$ is the relative humidity of the ambient air. On the dry part of the substrate the mass flux is zero, i.e., $\partial c / \partial n = 0$, where $\boldsymbol{n}$ is the unit normal. Assuming that the heat conduction and convection in the air can be neglected,[16,18-20] the local energy balance on the liquid-vapor interface is

$$k_L \nabla T \cdot \boldsymbol{n} = H_L D \nabla c \cdot \boldsymbol{n} \tag{2}$$

where $k_L$ is the thermal conductivity of the liquid, $H_L$ is the latent heat of evaporation, and $D$ is the diffusion coefficient of liquid vapor in the atmosphere.

### 3. Combined field approach

During drying, the heat conduction field inside the liquid droplet and the vapor diffusion field in the surrounding air interact with each other through the liquid-vapor interface: the vapor diffusion-controlled evaporation will lower the temperature of the liquid at the liquid surface, which can inversely affect the evaporative rate through its control on the saturation concentration of vapor right at the surface.[11-13,16] Thus, the liquid evaporation is a two-way coupling problem and numerical approaches have to be used to obtain the evaporation rate from the drying droplets.

From the above equations, it can be seen that both the two coupled fields in the liquid evaporation are governed by the Laplace's equation. This enables the possibility of combining these two fields into one single physics field which is also governed by the same differential equation. All we need to do is transfer the boundary conditions at the interface between the coupled fields (e.g., equations 1-2 in the text) to the boundary conditions at the interface between two media in one single field. This may be achieved through the nondimensionalization of the physical quantities of the fields.

Here, by choosing the scaling factor for the nondimensionalization of the temperature in the liquid droplet and that of the vapor concentration in the air as follows: $\tilde{T}_1 = \dfrac{b(T - T_0)}{c_0(1-H)}$, and $\tilde{T}_2 = \dfrac{c - c_0}{c_0(1-H)}$, the Laplace's equations and the boundary conditions in the droplet evaporation can be rewritten as

$$\nabla^2 \tilde{T}_1 = 0 \ \text{for } 0 \leq z \leq h(r),\ r \leq R \tag{3}$$

$$\nabla^2 \tilde{T}_2 = 0 \ \text{for } z \geq h(r),\ r \leq R;\ r > R \tag{4}$$

$$\tilde{T}_1 = \tilde{T}_2,\ \frac{\partial \tilde{T}_1}{\partial \boldsymbol{n}} = \boldsymbol{Ec} \frac{\partial \tilde{T}_2}{\partial \boldsymbol{n}} \ \text{for } z = h(r),\ r \leq R \tag{5}$$

$$\tilde{T}_1 = 0 \ \text{for } z = 0,\ r \leq R \tag{6}$$

$$\tilde{T}_2 = -1 \ \text{for } z = \infty,\ r = \infty \tag{7}$$

$$\frac{\partial \tilde{T}_2}{\partial \boldsymbol{n}} = 0 \ \text{for } z = 0,\ r > R \tag{8}$$

where $h(r)$ is the height of the droplet, and $\boldsymbol{Ec} = \dfrac{H_L D b}{k_L}$ is a dimensionless number.

If we consider $\tilde{T}_1$, $\tilde{T}_2$ as the temperature in the liquid domain and that in the air domain respectively, equations (3-8) just represent the heat conduction in both the droplet region with thermal conductivity of 1 and the air region with thermal conductivity of **Ec**. Thus, the coupled two fields have been combined into one single heat conduction field, and consequently, $\tilde{T}_1$ and $\tilde{T}_2$ can be numerically solved from equations (3-8) without any iteration between the different fields. Once $\tilde{T}_1$ and $\tilde{T}_2$ are known, the temperature in the droplet, the vapor concentration in the atmosphere, and the evaporation flux from the droplet surface can be easily computed.

It can also be seen that, for a droplet with a given contact angle, the above nondimensional equations for the droplet evaporation are governed only by the dimensionless number **Ec**. Its definition implies that the number **Ec** combines the effects of the thermal properties of the liquid, the atmospheric pressure, and the temperature dependence of the saturation concentration of vapor, and thus can be used to evaluate the strength of the evaporative cooling in the droplet evaporation.

## 4. Conclusions

By choosing the scaling factor for the nondimensionalization of the temperature in the liquid droplet and that of the vapor concentration in the air, the heat conduction field inside the liquid droplet and the vapor diffusion field in the surrounding air have been combined into one single "quasi-temperature" field. This makes the iteration between the coupled fields unneeded in solving numerically the two-way coupling problem. The present approach can be used to understand thoroughly the evaporative cooling effect in the liquid evaporation and may have more general applications to coupled field problems in which all the fields have the same governing equation.